\documentclass{article}
\usepackage{spconf,amsmath,graphicx,hyperref,amssymb,amsfonts}

\usepackage{textcomp}
\usepackage{xcolor}
\usepackage{tikz}
\usepackage{array}
\usepackage{multirow}
\usepackage{booktabs}
\usepackage{makecell}

\usepackage{subcaption}
\usepackage{adjustbox}

\title{A Stage-Wise Learning Strategy with Fixed Anchors for \\Robust Speaker Verification}
\name{Bin Gu\textsuperscript{1},
Lipeng Dai\textsuperscript{2},
Huipeng Du\textsuperscript{2},
Haitao Zhao\textsuperscript{1},
Jibo Wei\textsuperscript{1}}
\address{\textsuperscript{1}National University of Defense Technology, Changsha, China \\\textsuperscript{2}University of Science and Technology of China, Hefei, China}

\begin{document}

%
\maketitle

\begin{abstract}
Learning robust speaker representations under noisy conditions presents significant challenges, which requires careful handling of both discriminative and noise-invariant properties. In this work, we proposed an anchor-based stage-wise learning strategy for robust speaker representation learning. Specifically, our approach begins by training a base model to establish discriminative speaker boundaries, and then extract anchor embeddings from this model as stable references. Finally, a copy of the base model is fine-tuned on noisy inputs, regularized by enforcing proximity to their corresponding fixed anchor embeddings to preserve speaker identity under distortion. Experimental results suggest that this strategy offers advantages over conventional joint optimization, particularly in maintaining discrimination while improving noise robustness. The proposed method demonstrates consistent improvements across various noise conditions, potentially due to its ability to handle boundary stabilization and variation suppression separately.
\end{abstract}
\section{Introduction}
\label{sec:intro}
Speaker verification, which automatically determines whether two speech samples originate from the same person, has evolved significantly with recent technological advancements \cite{b1,b2,b3,b4,b5,b6}. However, their performance still degrades significantly when deployed in real-world environments with background noise, reverberation, and other acoustic distortions. This robustness gap stems from a fundamental challenge, in which SV requires learning features that are simultaneously discriminative (to distinguish between speakers) and invariant (to ignore non-speaker variations like noise). 

Current approaches to address this challenge can be broadly categorized by operating level. Feature-level methods typically incorporate speech enhancement modules to clean noisy inputs before extracting speaker characteristics \cite{b7,b8,b9,b10,b11,b13,b14}. Embedding-level methods instead focus on learning noise-invariant embeddings directly through advanced learning algorithms\cite{b16,b17,b15,b12,b18}. While both kinds of methods have shown promise, the embedding-level approach offers distinct advantages in terms of system compatibility and implementation simplicity. Our research contributes to this important direction by developing novel learning paradigms for noise-robust speaker representation.

Robust speaker representation learning has developed several effective methodologies to handle noisy environments. Disentanglement learning stands as one prominent solution, employing attribute decoupling techniques to extract noise-invariant speaker representations. The fundamental principle involves training networks to generate representations that confuse noise-type classifiers while preserving both speaker identity and spectral reconstruction capability. For example, \cite{b16} utilize explicit disentanglement of noise-sensitive and noise-invariant components to enhance speaker features robustness. Building on this foundation, \cite{b17} advanced the approach through adversarial training to make representations indistinguishable across different noise domains, demonstrating improved performance in challenging noisy datasets. Contrastive learning offers another powerful framework by directly optimizing the geometry of the speaker embedding space. These methods formulate the learning objective to simultaneously minimize distances between clean and noisy samples from the same speaker while maximizing separation between different speakers. In this paradigm, \cite{b15} jointly optimize speaker classification loss and either Euclidean or cosine distances between clean-noisy pairs. \cite{b12} developed a modified InfoNCE loss incorporating penalty terms and adaptive loss weight to better handle complex distribution relationships in noisy conditions. Beyond these established approaches, stable learning techniques have recently emerged to address dataset biases and improve generalization. These methods focus on identifying and eliminating spurious correlations in training data. For example, the work \cite{b18} showed how robust feature selection can enhance performance on unseen noise conditions. 

Although existing approaches have shown promising results, their effectiveness heavily relies on carefully balancing multiple loss functions through joint optimization. These methods typically require careful hyper-parameter tuning to achieve the delicate equilibrium between intra-class compactness and inter-class separation. The optimization process presents inherent challenges that excessive intra-class compression may lead to ambiguous decision boundaries between speakers, while over-emphasizing inter-class separation could prevent proper alignment of noisy and clean samples from the same speaker. This fundamental trade-off often results in suboptimal model performance \cite{b24,b25}, which complicates the training process.

\begin{figure*}[t]
\centerline{\includegraphics[width=0.95\linewidth]{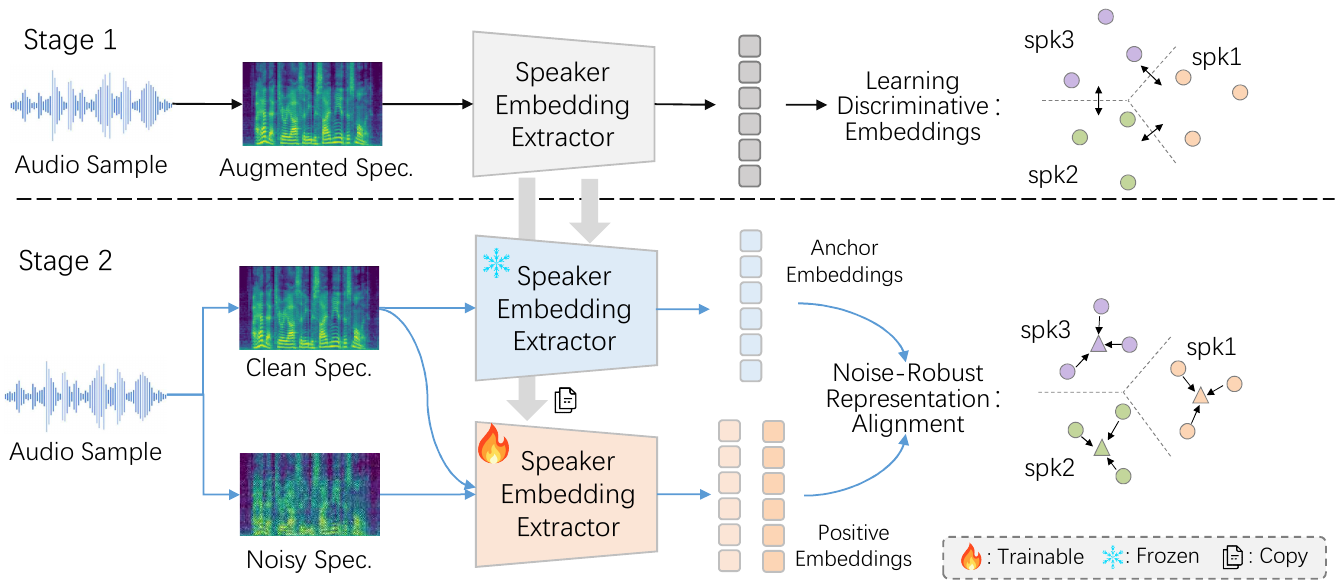}}
\caption{The stage-wise robust speaker representation learning framework.}
\label{fig1}
\vspace{-0.3cm}
\end{figure*}

To address these challenges, we propose a stage-wise robust feature learning method based on a fixed-anchor guidance framework and an anchor-driven intra-class variance suppression loss, which first establishes speaker discriminability and then enhancing noise robustness. Specifically, we train a base model which focuses on discriminating different speakers and freeze it to generate anchor embeddings from clean speech. Then, a trainable copy of the base model processes noisy inputs while being optimized to align with those fixed anchor embeddings. Experimental results on VoxCeleb1 demonstrate that our method achieves superior intra-class compactness and inter-class separation compared to joint training baselines, while outperforming existing approaches in terms of overall performance.

\section{Proposed Method}

\subsection{Overview of the Framework}

The proposed framework, illustrated in Fig.~\ref{fig1}, follows a two-stage learning procedure. In the first stage, the model adopts a standard training recipe where the feature extractor $g(\cdot)$ is optimized through speaker classification loss to learning discriminative embeddings. The objective can be formulated as 
\begin{equation}
\mathcal{L}_1 = -\log p(y|\mathbf{x})
\end{equation}
where $p(y|\mathbf{x})$ represents the predicted probability for input $\mathbf{x}$ belonging to speaker $y$, typically implemented using softmax or its variants.

Then, the second stage begins by duplicating the trained extractor $g(\cdot)$ into a fixed anchor branch $g_f(\cdot)$ that processes clean samples, and a trainable branch $g_t(\cdot)$ that handles clean and noisy samples. The optimization then minimizes the divergence between corresponding embeddings through the following objective:
\begin{equation}
\mathcal{L}_2 = D(\mathbf{x}_{a}, \mathbf{x}_{p})
\end{equation}
where $D(\cdot,\cdot)$ measures the distance between embeddings,  $\mathbf{x}_{a}$ and $\mathbf{x}_{p}$ denote anchor and positive samples from the same utterance. The final optimized $g_t(\mathbf{x})$ is deployed as the speaker embedding extractor during inference.

 This staged approach effectively preserves the inter-speaker discriminability learned in the first stage while enhancing robustness against noise. The fixed anchor embeddings serve as stable reference points in the high-dimensional space, preventing excessive drift of decision boundaries caused by noisy samples. Consequently, this strategy mitigates the common issue of blurred inter-speaker boundaries that often occurs when aggressively compressing intra-class variations, thereby maintaining clear discrimination between different speakers while improving noise robustness.

\begin{table*}[t]
\small
\centering
\caption{COMPARISON RESULTS (EER\%) ON VOXCELEB1 TEST SET WITH MUSAN DATA AT VARIOUS SNRS}
\vspace{-0.2cm}
\label{tab:results}
\resizebox{\textwidth}{!}{
\begin{tabular}{c|c|ccccccccccc}
\cline{1-13}
\toprule
\multicolumn{2}{c|}{Training Set}&\multicolumn{11}{c}{VoxCeleb1}\\

\cline{1-13}
Noise Type& SNR &Baseline& VoiceID\textsuperscript{\cite{b8}}  & FSEF\textsuperscript{\cite{b9}}  & NDML\textsuperscript{\cite{b16}}  & WSVIL\textsuperscript{\cite{b15}}  & ExU-Net\textsuperscript{\cite{b10}}  & SEU-Net\textsuperscript{\cite{b18}}& Diff-SV\textsuperscript{\cite{b11}}  & NDAL\textsuperscript{\cite{b17}}  &NISRL\textsuperscript{\cite{b12}}  & Proposed \\
\cline{1-13}

\multicolumn{2}{c|}{Original Set} & 1.98 & 6.79 & 4.26 & 2.90 & 3.12 & 2.76& 2.52 & 2.35 & 2.63 & 2.40 & \textbf{1.88} \\
\cline{1-13}
\multirow{5}{*}{Babble} 
& 0 & 9.30 & 38.0 & 27.6 & 11.0 & 11.8 & 9.57 &8.54 & 8.74 & \textbf{6.43} & 7.81 &7.40\\
& 5 & 4.56 & 27.1 & 15.3 & 6.13 & 5.97 & 5.52 &5.16 & 4.51 & 4.44 & 4.25 &\textbf{3.67}\\
& 10 & 2.99 & 16.7 & 9.04 & 4.28 & 4.44 & 4.06 &3.67 & 3.33 & 3.59 & 3.28 &\textbf{2.46}\\
& 15 & 2.45 & 11.3 & 6.47 & 3.52 & 3.73 & 3.28 &3.10 & 2.82 & 3.08 & 2.78 &\textbf{2.13}\\
& 20 & 2.18 & 8.99 & 5.41 & 3.21 & 3.36 & 2.99 &2.79 & 2.61 & 2.87 & 2.60 &\textbf{1.95}\\
\cline{1-13}

\multirow{5}{*}{Music}
& 0 & 5.82 & 16.2 & 8.47 & 10.8 & 7.79 & 7.35 &6.25 & 6.04 & 5.87 & 5.19 &\textbf{4.51}\\
& 5 & 3.57 & 11.4 & 6.31 & 6.52 & 5.23 & 4.90 &4.36 & 3.96 & 4.19 & 3.58 &\textbf{2.95}\\
& 10 & 2.73 & 9.13 & 5.14 & 4.66 & 4.11 & 3.69 &3.55 & 3.10 & 3.53 & 3.11 &\textbf{2.35}\\
& 15 & 2.28 & 8.10 & 4.71 & 3.67 & 3.63 & 3.14  &3.10 & 2.75 & 3.23& 2.75 &\textbf{2.03}\\
& 20 & 2.13 & 7.48 & 4.56 & 3.21 & 3.30 & 2.93 &2.79 & 2.60 & 3.09 & 2.57 &\textbf{1.92}\\
\cline{1-13}

\multirow{5}{*}{Noise}
& 0 & 7.3 & 16.6 & 7.88 & 10.2 & 7.34 & 6.80 &6.41 & 6.01 & 6.14 & \textbf{4.94} &5.29\\
& 5 & 4.45 & 12.3 & 6.42 & 6.96 & 5.65 & 5.23 &4.42 & 4.52 & 4.00 & 3.69 &\textbf{3.52}\\
& 10 & 3.14 & 9.86 & 5.50 & 5.02 & 4.35 & 4.07 &3.74 & 3.49 & 3.23 & 3.43 &\textbf{2.60}\\
& 15 & 2.57 & 8.69 & 4.87 & 3.91 & 3.85 & 3.39 &3.20 & 2.93 & 2.97 & 2.94 &\textbf{2.27}\\
& 20 & 2.25 & 7.83 & 4.66 & 3.40 & 3.44 & 3.10 &2.92 & 2.64 & 2.80 & 2.68 &\textbf{1.96}\\
\cline{1-13}

\multicolumn{2}{c|}{Average} & 3.73& 13.5 & 7.91 & 5.59 & 5.07 & 4.55 &4.16  & 3.90 & 3.88 & 3.62 & \textbf{3.05}\\
\bottomrule
\end{tabular}
}
\label{tab1}
\vspace{-0.5cm}
\end{table*}

\subsection{Anchor-Driven Intra-Variance Suppression}

The proposed method optimizes the embedding distance by minimizing anchor-positive pairs between two parallel extractors. Based on the second-stage $\mathcal{L}_2$ optimization for intra-class variance described earlier, we specifically employ an exponential form of cosine distance to measure divergence:
\begin{equation}
\begin{aligned}
\mathcal{L}_2 = &K(\mathbf{x}_{clean}, \mathbf{x}_{noise}) + K(\mathbf{x}_{clean}, \mathbf{x}_{clean})-\log p(y|\mathbf{x}_{noise}), \\
&K(\mathbf{x}_1, \mathbf{x}_2) = \exp\left(m \cdot (1 - \cos(g_f(\mathbf{x}_1), g_t(\mathbf{x}_2)))\right)
\end{aligned}
\end{equation}
where $m$ is a scaling factor. The cosine distance is chosen over Euclidean distance because it better captures the angular divergence between speaker embeddings, while the exponential term amplifies the loss gradient to accelerate convergence. Crucially, the $\mathcal{L}_2$ also minimizes the distance between clean-sample embeddings from both extractors, which serves as a regularization term to prevent significant deviation of clean samples when noisy samples converge toward their anchors. In addition, a classification loss is jointly applied during the second stage, serving as an extra regularization to preserve inter-speaker discriminability and avoid collapse of class separability among noisy samples.

This approach differs from mainstream methods that minimize distances between positive-negative pairs extracted from the same trainable extractor. Since $g_f(\mathbf{x})$ is frozen, the optimization process receives more stable learning signals. This design avoids the oscillatory behavior that occurs when both embedding vectors are dynamically updated, thereby effectively preventing model collapse. The frozen anchor branch maintains stable reference points in the embedding space, while the trainable branch learns to produce noise-robust representations that remain properly aligned with the clean-speech topology.

\section{Experiments}

\subsection{Data}
We evaluate our system on the VoxCeleb1 dataset \cite{b20}, using its standard development set with 1,211 speakers for training and test set containing 37,720 trials from 40 speakers. To thoroughly assess robustness, we create noisy evaluation conditions by mixing clean utterances with noise samples from both MUSAN \cite{b21} and Nonspeech100 \cite{b22} at signal-to-noise ratios ranging from 0 dB to 20 dB in 5 dB increments. MUSAN provides three noise categories including babble, music and environmental noise. Following the protocol in \cite{b11}, we strictly separate these noise samples into non-overlapping training and testing subsets to prevent data leakage. During model training, we apply online data augmentation through additive noise mixing with randomly selected training-set noise samples combined with convolutional reverberation using simulated room impulse responses. All remaining noise samples are reserved exclusively for constructing evaluation sets that cover both in-domain and out-of-domain noise conditions.

\subsection{Implementation Details}

The baseline system extracts 80-dimensional log-mel filterbank as acoustic features and uses a ResNet34 backbone with 32 initial channels to generate 256-dimensional speaker embeddings via statistics pooling. The model is trained using SGD optimizer with AAM-Softmax loss function, employing mixed-precision FP16 training on two NVIDIA RTX 5070 Ti GPUs with a per-GPU batch size of 128. Each utterance is randomly augmented with either additive noise (MUSAN samples at 0-20 dB SNR) or convolutional reverberation (simulated room impulse responses). For the proposed system, both training stages maintain the same configurations with those in baseline, including learning rate, number of epoches, and data augmentation pipeline. The scaling factor $m$ in Eq. 3 is empirically set to 5. Our implementation builds upon the Wespeaker toolkit \cite{b23}, which provides standardized configurations for speaker recognition systems. The toolkit handles essential training components including gradient synchronization and learning rate scheduling. For complete implementation details regarding the network architecture and training procedures, readers may refer to the official Wespeaker documentation.

\begin{figure*}[t]
    \centering
    \begin{subfigure}[b]{0.32\textwidth}
        \includegraphics[width=\linewidth]{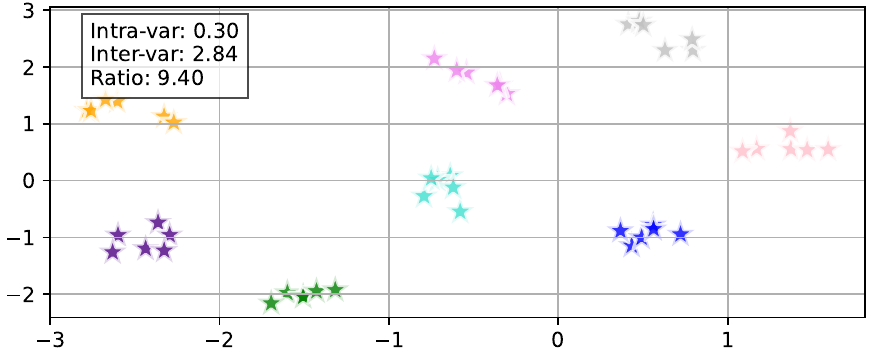}
        \caption{}
        \label{fig:a}
    \end{subfigure}
    \hfill
    \begin{subfigure}[b]{0.32\textwidth}
        \includegraphics[width=\linewidth]{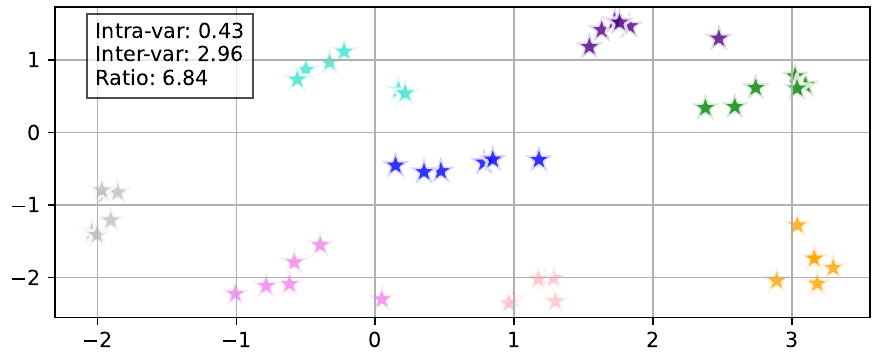}
        \caption{}
        \label{fig:b}
    \end{subfigure}
    \hfill
    \begin{subfigure}[b]{0.32\textwidth}
        \includegraphics[width=\linewidth]{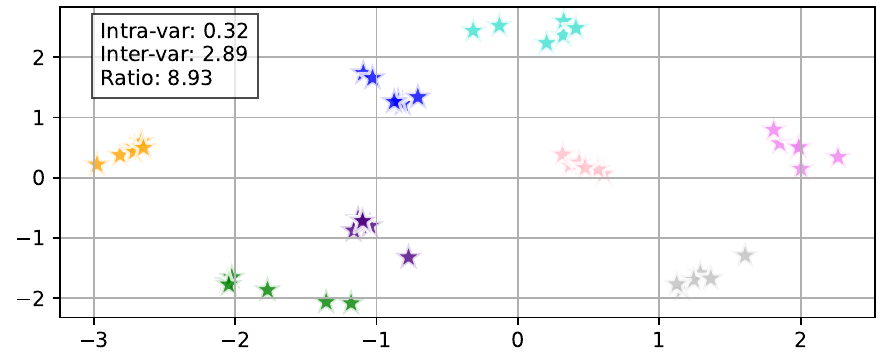}
        \caption{}
        \label{fig:c}
    \end{subfigure}
    \\
    \centering
    \begin{subfigure}[b]{0.325\textwidth}
        \includegraphics[width=\linewidth]{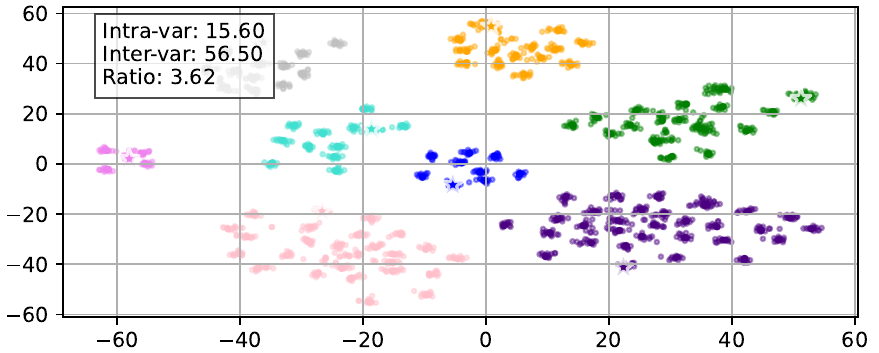}
        \caption{}
        \label{fig:d}
    \end{subfigure}
    \hfill
    \begin{subfigure}[b]{0.325\textwidth}
        \includegraphics[width=\linewidth]{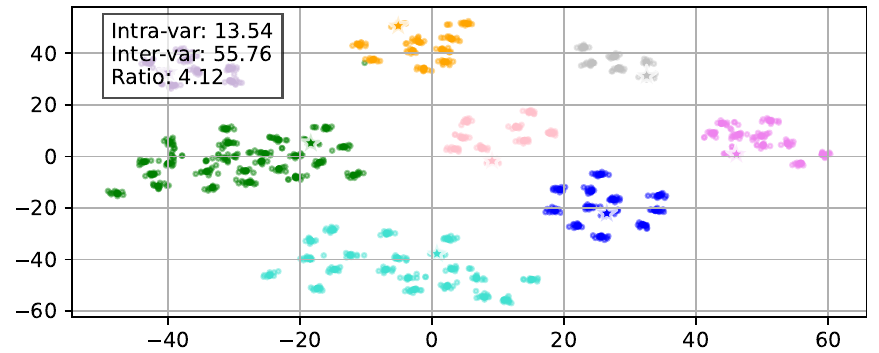}
        \caption{}
        \label{fig:e}
    \end{subfigure}
    \hfill
    \begin{subfigure}[b]{0.325\textwidth}
        \includegraphics[width=\linewidth]{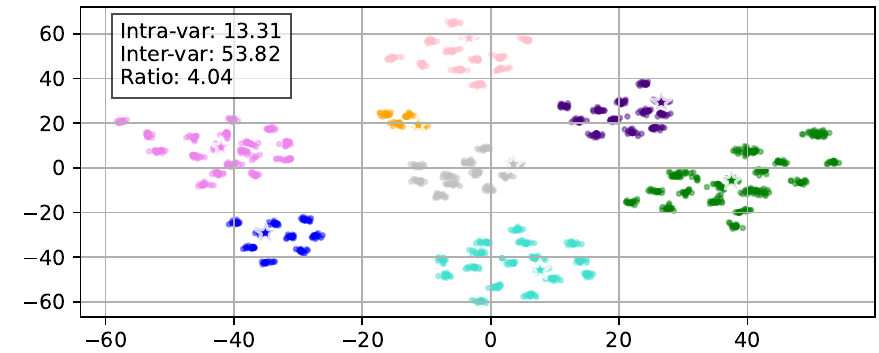}
        \caption{}
        \label{fig:f}
    \end{subfigure}

    \caption{The t-SNE visualization of speaker embeddings of the test set. Colors represent different speakers, with stars and circles denoting clean and noisy samples respectively. Each subplot's legend shows inter-class variance (between-speaker separation), intra-class variance (within-speaker consistency), and their ratio (higher values indicate better discriminability). Subfigures (a)-(c) visualize speaker embeddings from clean samples, while (d)-(f) display corresponding noisy-sample embeddings from the same utterances. Systems compared are: (a,d) baseline, (b,e) joint-learning, and (c,f) our proposed system}
    \label{fig:combined}
    \vspace{-0.3cm}
\end{figure*}

\subsection{Results}

As shown in Table \ref{tab1}, our proposed method demonstrates significant improvements over both the baseline and existing techniques, achieving the best overall performance. Notably, our baseline system outperforms most competing methods in the in-domain test scenario (i.e., babble, music, and noise conditions), which we attribute to the more advanced model learning configuration of the Wespeaker Toolkit. Moreover, compared to contrastive learning or disentanglement learning-based methods (NISRL \& NDAL), our feature learning approach exhibits superior effectiveness. For the out-of-domain tests shown in Table \ref{tab2}, the proposed method obtain a marked performance enhancement under unseen noise conditions, confirming its strong generalization capability in unknown scenarios. 

\subsection{Ablation Study and Analysis}
\begin{table}[t]
\centering
\small
\caption{COMPARISON RESULTS (EER\%) ON VOXCELEB1 TEST SET WITH NONSPEECH100 DATA AT VARIOUS SNRs.}
\begin{tabular}{c|cccccc}
\toprule 
{SNR}& {Baseline} & {SEU-Net} & {Diff-sv} & {NDAL} & {NISRL} &{Proposed} \\
\cline{1-7}
0 & 10.17 & \textbf{5.99} & 8.23 & 7.57 & 6.41 &6.85 \\
5 & 5.53 & 4.58 & 5.06 & 5.49 & 4.57 &\textbf{4.13}\\
10 & 3.79 & 3.74 & 3.85 & 4.03 & 3.55 &\textbf{2.98}\\
15 & 2.74 & 3.15 & 3.19 & 3.36 & 2.99 &\textbf{2.25}\\
20 & 2.36 & 2.87 & 2.89 & 2.99 & 2.75 &\textbf{2.02}\\
\cline{1-7}
Average & 4.92 & 4.07 & 4.65 & 4.97 & 4.05 & \textbf{3.64}\\
\bottomrule
\end{tabular}
\label{tab2}
\vspace{-0.5pt}
\end{table}

As shown in Table \ref{tab3}, we compared results of the systems with joint learning or stage-wise optimization. During join-learning, system was trained from random initialization with a combined speaker classification loss and intra-variance suppression loss, and both anchor and positive vectors were extracted from the same trainable model. The experimental results reveal that the jointly trained model exhibits significant performance degradation compared to the baseline under clean test conditions, yet achieves superior performance in noisy environments. This suggests that while joint optimization may reduce inter-class discriminability, it enhances robustness under noisy testing conditions, potentially due to excessive intra-class compression leading to blurred decision boundaries. In contrast, our proposed method demonstrates consistently better performance across both clean and noisy scenarios, validating its effectiveness in maintaining discriminative power while improving robustness. 

Fig. 2 visually compares speaker embeddings across three systems (left to right: baseline, joint-learning, and proposed method). For clean samples (a-c), the legend reveals the joint-training system achieves poorer speaker discriminability (lower ratio) while our method maintains comparable inter-class boundaries to the baseline. In noisy conditions (d-f), the baseline shows significant degradation in clean-noisy sample consistency (particularly evident in dark purple/light pink clusters), whereas our method effectively suppresses noise-induced intra-class dispersion. These visualizations collectively demonstrate that our proposed approach achieves superior balance between inter-class separation and intra-class compactness under both clean and noisy conditions.

\begin{table}[t]
\centering
\small
\caption{COMPARISON RESULTS (AVERAGE EER\% ACROSS 5 SNR LEVELS) OF DIFFERENT SYSTEMS UNDER VARIOUS SYNTHETIC NOISE CONDITIONS.}
\begin{tabular}{c|cccc}
\toprule
Noise Type& Baseline & Join-Learning & Proposed \\
\cline{1-4}
Original & 1.98 & 2.22 & 1.88 \\
\cline{1-4}
Babble & 4.30 & 4.06 & 3.52 \\
Music & 3.31 & 3.03 & 2.75\\
Noise & 3.94 & 3.51 & 3.13\\
Nonspeech & 4.92 & 4.08 & 3.64\\
\cline{1-4}
Average & 3.69 & 3.38 & 2.98\\
\bottomrule
\end{tabular}
\label{tab3}
\end{table}

\section{Conclusion}

To address robust speaker verification in noisy environments, we propose a two-stage representation learning framework that first emphasizes inter-class discriminative optimization, then employs an anchor-driven intra-class variance suppression loss to enhance cosine similarity between clean-noisy sample pairs while constraining inter-class boundary fluctuations within a limited range. Experimental results demonstrate that our approach effectively improves model robustness while preserving intrinsic discriminative power. Visualization analyses further reveal that the method successfully mitigates the inherent tension between intra-class compactness and inter-class separability, achieving superior system performance compared to conventional approaches. The proposed technique's dual-phase optimization strategy is shown to maintain stable decision boundaries under varying noise conditions while promoting more concentrated feature distributions within speaker classes.

\bibliographystyle{IEEEbib}
\bibliography{strings,refs}

@article{b1,
  title={Speaker recognition by machines and humans: A tutorial review},
  author={Hansen, John HL and Hasan, Taufiq},
  journal={IEEE Signal Process. Mag.},
  volume={32},
  pages={74--99},
  year={2015},
  publisher={IEEE}
}

@inproceedings{b2,
  title={X-vectors: Robust dnn embeddings for speaker recognition},
  author={Snyder, David and Garcia-Romero, Daniel and Sell, Gregory and Povey, Daniel and Khudanpur, Sanjeev},
  booktitle={Proc. IEEE Int. Conf. Acoustics, Speech, and Signal Processing (ICASSP)},
  pages={5329--5333},
  year={2018}
}

@article{b3,
  title={Overview of speaker modeling and its applications: From the lens of deep speaker representation learning},
  author={Wang, Shuai and Chen, Zhengyang and Lee, Kong Aik and Qian, Yanmin and Li, Haizhou},
  journal={ IEEE/ACM Trans. Audio, Speech Lang. Processing},
  year={2024},
  volume={32},
  pages={4971--4998},
  publisher={IEEE}
}

@inproceedings{b5,
  title={Ecapa-tdnn: Emphasized channel attention, propagation and aggregation in tdnn based speaker verification},
  author={Desplanques, Brecht and Thienpondt, Jenthe and Demuynck, Kris},
  booktitle={Proc. Interspeech},
  page={3830--3834},
  year={2020}
}

@article{b4,
  title={Memory storable network based feature aggregation for speaker representation learning},
  author={Gu, Bin and Guo, Wu and Zhang, Jie},
  journal={IEEE/ACM Trans. Audio, Speech Lang. Processing},
  volume={31},
  pages={643--655},
  year={2023},
  publisher={IEEE}
}

@article{b6,
  title={A dynamic convolution framework for session-independent speaker embedding learning},
  author={Gu, Bin and Zhang, Jie and Guo, Wu},
  journal={IEEE/ACM Trans. Audio, Speech Lang. Processing},
  volume={31},
  pages={3647--3658},
  year={2023},
  publisher={IEEE}
}

@inproceedings{b7,
  title={Feature enhancement with deep feature losses for speaker verification},
  author={Kataria, Saurabh and Nidadavolu, Phani Sankar and Villalba, Jes{\'u}s and Chen, Nanxin and Garcia-Perera, Paola and Dehak, Najim},
  booktitle={Proc. IEEE Int. Conf. Acoustics, Speech, and Signal Processing (ICASSP)},
  pages={7584--7588},
  year={2020}
}

@inproceedings{b8,
  title={VoiceID Loss: Speech Enhancement for Speaker Verification},
  author={Shon, Suwon and Tang, Hao and Glass, James},
  booktitle={Proc. Interspeech},
  pages={2888--2892},
  year={2019}
}

@article{b9,
  title={A fused speech enhancement framework for robust speaker verification},
  author={Wu, Yanfeng and Li, Taihao and Zhao, Junan and Wang, Qirui and Xu, Jing},
  journal={IEEE Signal Processing Letters},
  volume={30},
  pages={883--887},
  year={2023}
}

@inproceedings{b10,
  title={Extended U-Net for Speaker Verification in Noisy Environments},
  author={Kim, Ju Ho and Heo, Jungwoo and Shim, Hye Jin and Yu, Ha Jin},
  booktitle={Proc. Interspeech},
  pages={590--594},
  year={2022}
}

@inproceedings{b11,
  title={Diff-sv: A unified hierarchical framework for noise-robust speaker verification using score-based diffusion probabilistic models},
  author={Kim, Ju-ho and Heo, Jungwoo and Shin, Hyun-seo and Lim, Chan-yeong and Yu, Ha-Jin},
  booktitle={Proc. IEEE Int. Conf. Acoustics, Speech, and Signal Processing (ICASSP)},
  pages={10341--10345},
  year={2024}
}

@inproceedings{b12,
  title={Aligning Noisy-Clean Speech Pairs at Feature and Embedding Levels for Learning Noise-Invariant Speaker Representations},
  author={Li, Zuoliang and Ai, Yang and Zhang, Jie and Peng, Shengyu and Guan, Yu and Gu, Bin and Guo, Wu},
  booktitle={Proc. IEEE Int. Conf. Acoustics, Speech, and Signal Processing (ICASSP)},
  pages={1--5},
  year={2025}
}

@inproceedings{b13,
  title={Audio enhancing with DNN autoencoder for speaker recognition},
  author={Plchot, Oldrich and Burget, Lukas and Aronowitz, Hagai and Matejka, Pavel},
  booktitle={Proc. IEEE Int. Conf. Acoustics, Speech, and Signal Processing (ICASSP)},
  pages={5090--5094},
  year={2016}
}

@article{b14,
  title={Front-end speech enhancement for commercial speaker verification systems},
  author={Eskimez, Sefik Emre and Soufleris, Peter and Duan, Zhiyao and Heinzelman, Wendi},
  journal={Speech Communication},
  volume={99},
  pages={101--113},
  year={2018},
  publisher={Elsevier}
}

@inproceedings{b15,
  title={Within-sample variability-invariant loss for robust speaker recognition under noisy environments},
  author={Cai, Danwei and Cai, Weicheng and Li, Ming},
  booktitle={Proc. IEEE Int. Conf. Acoustics, Speech, and Signal Processing (ICASSP)},
  pages={6469--6473},
  year={2020}
}

@inproceedings{b16,
  title={Noise-disentanglement metric learning for robust speaker verification},
  author={Sun, Yao and Zhang, Hanyi and Wang, Longbiao and Lee, Kong Aik and Liu, Meng and Dang, Jianwu},
  booktitle={Proc. IEEE Int. Conf. Acoustics, Speech, and Signal Processing (ICASSP)},
  pages={1--5},
  year={2023}
}

@inproceedings{b17,
  title={A Joint Noise Disentanglement and Adversarial Training Framework for Robust Speaker Verification},
  author={Xing, Xujiang and Xu, Mingxing and Zheng, Thomas Fang},
  booktitle={Proc. Interspeech},
  pages={707--711},
  year={2024}
}

@inproceedings{b18,
  title={Stable Extended U-Net for Noise-Robust Speaker Verification},
  author={Wang, Zonghui and Fang, Zhihua and He, Liang},
  booktitle={Proc. IEEE Int. Conf. Acoustics, Speech, and Signal Processing (ICASSP)},
  pages={1--5},
  year={2025}
}

@inproceedings{b20,
  title={VoxCeleb: a large-scale speaker identification dataset},
  author={Nagrani, Arsha and Chung, Joon Son and Zisserman, Andrew},
  booktitle={Proc. Interspeech},
  year={2017},
  volume={3},
  pages={2616--2620}
}

@article{b21,
  title={Musan: A music, speech, and noise corpus},
  author={Snyder, David and Chen, Guoguo and Povey, Daniel},
  journal={arXiv preprint arXiv:1510.08484},
  year={2015}
}

@article{b22,
  title={A tandem algorithm for pitch estimation and voiced speech segregation},
  author={Hu, Guoning and Wang, DeLiang},
  journal={IEEE Trans. Audio, Speech, and Lang. Processing},
  volume={18},
  pages={2067--2079},
  year={2010},
  publisher={IEEE}
}

@inproceedings{b23,
  title={Wespeaker: A research and production oriented speaker embedding learning toolkit},
  author={Wang, Hongji and Liang, Chengdong and Wang, Shuai and Chen, Zhengyang and Zhang, Binbin and Xiang, Xu and Deng, Yanlei and Qian, Yanmin},
  booktitle={Proc. IEEE Int. Conf. Acoustics, Speech, and Signal Processing (ICASSP)},
  pages={1--5},
  year={2023}
}

@inproceedings{
b24,
title={{\{}GS{\}}hard: Scaling Giant Models with Conditional Computation and Automatic Sharding},
author={Dmitry Lepikhin and HyoukJoong Lee and Yuanzhong Xu and Dehao Chen and Orhan Firat and Yanping Huang and Maxim Krikun and Noam Shazeer and Zhifeng Chen},
booktitle={Proc. Inter. Conf. Learning Representations (ICLR)},
year={2021},
url={https://openreview.net/forum?id=qrwe7XHTmYb}
}

@inproceedings{b25,
  title={Gradient surgery for multi-task learning},
  author={Yu, Tianhe and Kumar, Saurabh and Gupta, Abhishek and Levine, Sergey and Hausman, Karol and Finn, Chelsea},
  booktitle={Proc. Advances in Neural Information Processing Systems (NeurIPS)},
  pages={5824--5836},
  year={2020}
}

\end{document}